\documentclass[aps,pra,showpacs,superscriptaddress,10pt,longbibliography]{revtex4-1}
\usepackage[T1]{fontenc}

\pdfoutput=1
\usepackage[dvipsnames,rgb,dvips,hyperref]{xcolor}
\usepackage{graphicx}
\usepackage{float}
\usepackage{overpic}
\usepackage{amsmath,amssymb,amsfonts,bbm}
\usepackage{hyperref}
\definecolor{linkblue}{HTML}{2200CC}
\hypersetup{
    colorlinks,
    linkcolor=linkblue,
    citecolor=linkblue,
    urlcolor=linkblue
}

\makeatletter
\def\amsbb{\use@mathgroup \M@U \symAMSb}
\makeatother
\usepackage[bbgreekl]{mathbbol}
\usepackage{mathrsfs}

\newcommand{\lc}{\ensuremath{\varepsilon}}

\newcommand{\fourier}[1]{\ensuremath{\mathcal F #1}}
\newcommand{\invfourier}[1]{\ensuremath{\mathcal F^{-1} #1}}

\newcommand{\Tr}{\ensuremath{\textrm{Tr}\,}}

\newcommand\nn{\nonumber}
\newcommand\ve[1]{\boldsymbol{#1}}
\newcommand{\ma}[1]{\ensuremath{\amsbb{#1}}}
\newcommand{\rd}{\ensuremath{\textrm{d}}}
\newcommand{\tr}{\ensuremath{^{\rm T}}}

\newcommand{\Reyp}{\ensuremath{\textrm{Re}_{\rm p}}}
\newcommand{\De}{\ensuremath{\textrm{De}}}
\newcommand{\Wi}{\ensuremath{\textrm{Wi}}}

\newcommand{\eqnlab}[1]{\label{eqn:#1}}
\newcommand{\figlab}[1]{\label{fig:#1}}

\newcommand{\eqnref}[1]{(\ref{eqn:#1})}

\newcommand{\Eqnref}[1]{Eq.~(\ref{eqn:#1})}
\newcommand{\Figref}[1]{Fig.~\ref{fig:#1}}

\newcommand{\applab}[1]{\label{app:#1}}

\newcommand{\seclab}[1]{\label{sec:#1}}

\newcommand{\Appref}[1]{Appendix~\ref{app:#1}}

\newcommand{\Secref}[1]{Section~\ref{sec:#1}}

\begin{document}

\title{Spherical particle sedimenting in weakly viscoelastic shear flow}

\author{Jonas Einarsson}
\author{Bernhard Mehlig}
\affiliation{Gothenburg University, 412 96 Gothenburg, Sweden}
\date{\today}

\begin{abstract}
We consider the dynamics of a small spherical particle driven through an unbounded viscoelastic shear flow by an external force. We give analytical solutions to both the mobility problem (velocity of forced particle) and the resistance problem (force on fixed particle), valid to second order in the dimensionless Deborah and Weissenberg numbers, which represent the elastic relaxation time of the fluid relative to the rate of translation and the imposed shear rate. We find a shear-induced lift at $O(\Wi)$, a modified drag at $O(\De^2)$ and $O(\Wi^2)$, and a second lift that is orthogonal to the first, at $O(\Wi^2)$. The relative importance of these effects depends strongly on the orientation of the forcing relative to the shear. We discuss how these forces affect the terminal settling velocity in an inclined shear flow.
We also describe a new basis set of symmetric Cartesian tensors, and demonstrate how they enable general tensorial perturbation calculations such as the present theory. In particular this scheme allows us to write down a solution to the inhomogenous Stokes equations, required by the perturbation expansion, by a sequence of algebraic manipulations well suited to computer implementation.
\end{abstract}

\maketitle

\section{Introduction}

In this paper we consider the mobility of a small spherical particle driven through an unbounded viscoelastic shear flow by an external force. In a Newtonian fluid the velocity of the particle is determined by the balance between the Stokes drag and the external force, and it is unaffected by the shear flow because of the linearity of Stokes equations. But in a viscoelastic fluid the disturbance flow around the particle interacts non-linearly with the shear flow to induce viscoelastic stresses. As a consequence the mobility depends non-linearly on the forcing and the shear flow.

A viscoelastic shear flow can reduce the terminal velocity of a sphere when the applied shear flow is perpendicular to gravity \cite{van_den_brule_effects_1993,housiadas_drag_2012,padhy_simulations_2013,davino2015}. This so-called cross-shear flow is a model system for transport of particles in vertical cracks induced by hydraulic fracturing \cite{barbati_complex_2016}. Experiments by \citet{van_den_brule_effects_1993} first demonstrated that a cross-shear flow strongly reduces the settling velocity, and that fluid elasticity is the dominant mechanism. Recently, numerical simulations by \citet{padhy_simulations_2013,padhy_effect_2013} verified an increased drag on a sphere translating through a cross-shear flow, and showed that the experimental observation is explained by a combination of viscoelasticity and the effects of the nearby walls in the experiment. Calculations by \citet{housiadas_drag_2012,housiadas_rheological_2014} demonstrate that the drag is increased also in an unbounded viscoelastic cross-shear flow.

These studies concern settling in a cross-shear flow in which the gravity acts along the vorticity axis. In this case the physical system is invariant under a 180$^\circ$ rotation around the vorticity axis. This symmetry was exploited in both the analytical and numerical calculations to reduce the number of variables \cite{housiadas_drag_2012,housiadas_rheological_2014,padhy_simulations_2013}. In particular the only relevant force is the drag force, and the particle only rotates around the vorticity axis. The effect of the shear on settling is substantial in this symmetrical case, and this fact raises new questions: How are the dynamics affected when gravity acts at an angle to the vorticity? Are there additional forces and torques when the symmetry is broken? How does the drag change as the angle between gravity and flow vorticity changes?

In this paper we calculate the effect of an unbounded shear flow on the terminal particle velocity for any orientation of the external force relative to the shear. We must consequently abandon the simplifications of the symmetrical case. We must allow for lift forces in the flow-shear plane, and for rotation around any axis. Further, since viscoelasticity is a non-linear effect, it is not possible to construct the general result as a linear combination of results for two independent directions. Therefore we must solve the general perturbation problem for the flow velocity $\ve u$ and viscoelastic stress tensor $\ve \Pi$ around a translating and rotating particle in a shear flow. Our calculation relies on a perturbation theory for weak elasticity, valid to second order in the Deborah and Weissenberg numbers. These dimensionless numbers relate the elastic relaxation time of the fluid relative to the rate of translation and the shear rate. 

\citet{brunn1977,brunn1977a} and \citet{vishnampet2012} considered the first order of this problem, and both found lateral migration, although their detailed results do not agree with each other. The first part of our calculation is an independent check of their results, which we return to in \Secref{mobility}. \citet{housiadas2011a} and \citet{davino2008} considered the angular velocity of the sphere in the absence of an external force, and \citet{leslie1961} calculated the drag on a sphere in absence of shear flow. These two results coincide with our theory in their respective limits.

We solve the problem in tensorial form. Our solution does not refer to any coordinate representation such as spherical coordinates. Since the governing equations and boundary conditions are tensorial in nature, this substantially simplifies the calculations. All steps of the calculation are algebraic, and therefore well suited to computer implementation. To achieve this we introduce a new basis set of symmetric, rank-$n$ Cartesian tensors. We describe these tensors and how to calculate with them in some detail in \Secref{Ttensors}, because we expect that they will be useful for treating other problems too.

We present two related calculations. The first is the mobility problem, where we impose an external force $\ve F^{\mathrm{ext}}$ on the particle, and compute the resulting particle velocity $\ve v(\ve F^{\mathrm{ext}})$. This corresponds directly to the experimental protocol of for example \citet{van_den_brule_effects_1993}, where they release a sphere in a cylindrical Couette device and measure the steady settling velocity. The other question is the resistance problem, where we prescribe the particle velocity $\ve v$, and compute the resulting force $\ve F(\ve v)$ exerted by the fluid on the particle. This approach corresponds to the calculations by \citet{housiadas_rheological_2014} and numerical simulations by \citet{padhy_effect_2013}. The two are related, because given the solution to the mobility problem $\ve v(\ve F^{\mathrm{ext}})$, and the solution of the resistance problem $\ve F(\ve v)$, it must hold that $\ve F(\ve v)=-\ve F^{\mathrm{ext}}$.
In this paper we solve both the mobility problem and the resistance problem for a freely rotating spherical particle in an unbounded viscoelastic shear flow, with no restriction on the direction of $\ve v$ or $\ve F^{\mathrm{ext}}$ relative to the shear.

The rest of this paper is organized as follows. In \Secref{problem} we describe the problem, give the governing equations, and describe how we apply the Lorentz reciprocal theorem. In \Secref{Ttensors} we explain our algebraic solution of the inhomogeneous Stokes equation in terms of Cartesian tensors, and summarise their algebraic properties. We summarise our calculation and give the final result in \Secref{results}. We discuss the results and conclude in \Secref{discussion}.

\section{Problem formulation}\seclab{problem}

\subsection{Equation of motion and dimensionless parameters}

We consider the steady-state motion of a spherical particle of radius $a$, suspended in a viscoelastic fluid and subject to an external force ${\ve F}^{\mathrm{ext}}$. For concreteness we may think of the gravitational force ${\ve F}^{\mathrm{ext}} = 4\pi a^3(\rho_p-\rho_f)\ve g/3$. The particle moves with center-of-mass velocity ${\ve v}$, and rotates with angular velocity ${\ve \omega}$.
Far away from the particle the flow is a simple shear flow
\begin{align}
{\ve u}^\infty&={\ve \Omega}\times{\ve r} + {\ma S}{\ve r}\,,
\end{align}
where ${\ve \Omega}$ is half the flow vorticity, and the symmetric tensor ${\ma S}$ is the rate of strain.
In a simple shear flow the vorticity and strain are related by ${\ma S} {\ve \Omega}=0$ and $2|{\ve \Omega}|^2=\Tr {\ma S}{\ma S}$, in contrast to a general linear flow.

We work in dimensionless variables. The length scale is given by the particle radius $a$. The time scale is given by the reciprocal of the imposed shear rate $s=\sqrt{2\Tr {\ma S} {\ma S}}$, which also determines the scale of ${\ve u}^\infty$ to $sa$. The particle and disturbance flow velocities are nondimensionalized by the characteristic flow velocity $v_c$ past the particle. In the resistance problem, $v_c$ is simply the magnitude $|{\ve v}|$ of the imposed velocity ${\ve v}$. In the mobility problem we estimate the characteristic speed by $v_c={F}^\mathrm{ext}/a\mu$, related to the terminal velocity in Stokes flow under an external force of magnitude ${F}^\mathrm{ext}$. Here $\mu$ is the total viscosity, defined precisely in conjunction with the constitutive equations below. Stresses are made dimensionless by $v_c\mu/a$, and forces by $v_c \mu a$. In the remainder of this paper all quantities are dimensionless: $t'=s t$, $\ve r'={\ve r}/a$, $\ma S' = {\ma S}/s$, $\ve \Omega' = {\ve \Omega}/s$, $\ve F' = {\ve F}/(v_c\mu a)$, and so forth. We drop the primes since all quantities are dimensionless.

It follows that there are two dimensionless parameters that govern this problem, corresponding to the translational and rotational motion of the particle compared to the relaxation time $\lambda$ of the viscoelastic fluid. 
The Deborah number $\De = \lambda v_c/a$ is associated with the time scale of convective flow over the particle size. 
The Weissenberg number $\Wi=\lambda s$ is associated with the shear rate. The ratio $\alpha=\Wi/\De$ measures the relative importance of the imposed shear to the translational motion.

The perturbation theory in this paper is valid in the limit $\De\ll1$ and $\Wi\ll1$. This implies that the elastic part of the fluid relaxes quickly relative to the rate at which it is deformed by the moving particle and the shear flow. In the following we expand in $\De$, and treat the ratio $\alpha$ as an $O(1)$-quantity. But in the end we give the result in terms of $\De$ and $\Wi$. We note that the choice between $1/s$ and $a/v_c$ for the characteristic timescale is arbitrary up to factors of $\alpha = \Wi/\De$, which we assume to be $O(1)$.

We neglect the effects of fluid inertia. This requires that the viscous relaxation time of the fluid is shorter than the elastic relaxation time $\lambda$, so that we can neglect effects of inertio-elastic coupling. More precisely the particle Reynolds number $\Reyp=\rho_fv_ca/\mu\ll\De$. This condition is equivalent to $\rho_fa^2/\mu\ll\lambda$.

We write down the dimensionless governing equations for with respect to a frame moving with the steady center-of-mass velocity $\ve v$. In this frame the fluid pressure $p$ and velocity $\ve u$ satisfy
\begin{subequations}
\eqnlab{floweq1}
\begin{align}
  \nabla \cdot \ve \sigma&=0\,,\quad \ve \sigma = -p \ma I + (1-\mu_r) (\nabla \ve u+ (\nabla\ve u)\tr)+\mu_r\ve \Pi\,, \eqnlab{floweq1a}\\
  \nabla\cdot\ve u&=0\,.\eqnlab{floweq1b}
\end{align}
\end{subequations}
The viscoelastic stress tensor $\ve \Pi$ is modeled by the steady Oldroyd-B constituitive equations \cite{larson_constitutive_2013}. They describe a suspension of elastic dumbbells, which is one of the simplest models of an elastic polymer suspension that exhibits a normal stress difference in a shear flow. The equations are
\begin{align}
\ve \Pi + \De \left[(\ve u\cdot\nabla)\ve \Pi - (\nabla \ve u)\ve \Pi - \ve \Pi (\nabla \ve u)\tr\right]=\nabla \ve u + (\nabla \ve u)\tr\eqnlab{oldBdef}\,.
\end{align}
The parameter $\mu_r=\mu_p/(\mu_s+\mu_p)$ is the relative contribution to the total viscosity from the elastic polymers, relative to the solvent viscosity $\mu_s$. We denote the total viscosity $\mu=\mu_s+\mu_p$.

The flow problem in Eqns.~\eqnref{floweq1} and \eqnref{oldBdef} is completed by the no-slip boundary condition on the particle surface $S_p$ and that it approaches $\ve u^\infty$ as $|\ve r|\to\infty$:
\begin{align}
  \ve u &= \alpha \ve \omega \times \ve r,\quad\ve r \in S_p\,,\nn\\
  \ve u &\to  \alpha \ve u^\infty-\, \ve v\,,\quad|\ve r|\to \infty\,. \eqnlab{flowbc1}
\end{align}

The force and torque on the particle are given by
\begin{align}
  \ve F-\ve F^{\mathrm{ext}} &= \int_{S_p} \ve \sigma \cdot \ve {\rd S}\,,\\
  \ve T &= \int_{S_p} \ve r \times \ve \sigma \cdot \ve {\rd S}\,.\eqnlab{forceandtorque}
\end{align}
We do not consider any external torques in this paper and so $\ve T^\mathrm{ext}=0$.
We compute the forces on the particle, or the resulting velocities, with the Lorentz reciprocal theorem \cite{kim1991}. We outline how to apply it to our problem in the following Section.

\subsection{Reciprocal theorem}
In the context of perturbation theory, the Lorentz reciprocal theorem relates certain integral quantities such as the force or torque to order $n+1$, given the detailed flow solution to only order $n$ \cite{kim1991,leal1979}. For non-Newtonian flows in particular, the method has for example been used to calculate lateral migration \cite{ho_migration_1976}, and orbit drift of non-spherical particles \cite{leal1975}. In this Section we state the theorem as applicable to our problem and notation.

We denote the Newtonian part of the stress by $\ve \sigma^{N}=-p \ma I + (\nabla \ve u+ (\nabla\ve u)\tr)$, and the ``extra stress'' $\ve \sigma^{E}=\mu_r(\ve \Pi - (\nabla \ve u+ (\nabla\ve u)\tr))$, so that $\ve \sigma = \ve \sigma^{N}+\ve \sigma^{E}$.
The flow equation of motion \eqnref{floweq1a} is therefore
\begin{align}
 \nabla \cdot \ve \sigma^{N}=-\nabla \cdot \ve \sigma^{E}\,.   
\end{align}
The Lorentz reciprocal theorem for an arbitrary Stokes flow $(\ve{\tilde u},\ve{\tilde \sigma})$ and the flow defined in \Eqnref{floweq1} reads \cite{kim1991}
\begin{align}
  \int_S \ve {\tilde u} \cdot \ve\sigma^{N} \cdot \ve {\rd S}
  = \int_S \ve u \cdot \ve{\tilde\sigma} \cdot \ve {\rd S}
  + \int_V \ve{\tilde u}\cdot \nabla \cdot \ve\sigma^{N}{\rd V}\,.\eqnlab{rt01}
\end{align}
Here $V$ is any volume outside the particle, and $S$ denotes the surfaces bounding $V$. The vector $\ve {\rd S}=\ve n \rd S$, where $\ve n$ is the surface normal pointing out of $V$.
Using $\ve \sigma=\ve \sigma^{N}+\ve \sigma^{E}$, it follows from \eqnref{rt01} that
\begin{align}
  \int_S \ve {\tilde u} \cdot \ve\sigma \cdot \ve {\rd S}
  = 
  \int_S \ve u \cdot \ve{\tilde\sigma} \cdot \ve {\rd S}
  + \int_S \ve {\tilde u} \cdot \ve\sigma^{E} \cdot \ve {\rd S}
  - \int_V \ve{\tilde u}\cdot \nabla \cdot \ve\sigma^{E}{\rd V}\eqnlab{rt02}\,.
\end{align}
In the first two surface integrals in \eqnref{rt02} we identify the hydrodynamic force and torque on the particle, as given in \Eqnref{forceandtorque}.
We take $\ve{\tilde u}$ to be the Stokes flow around a spherical particle translating with velocity $\ve{\tilde v}$ and rotating with angular velocity $\ve{\tilde \omega}$ in an otherwise quiescent fluid. We write this auxiliary flow as $\ve{\tilde u}=\ma M_{\ve v}\ve{\tilde v}+\ma M_{\ve \omega}\ve{\tilde \omega}$, and we also know that $\ve{\tilde F}=-6\pi\tilde{\ve v}$ and $\ve{\tilde T}=-8\pi\tilde{\ve \omega}$.
Finally, upon taking the size of the volume $V$ to infinity, and inserting the boundary conditions \eqnref{flowbc1} into \Eqnref{rt02}, we find
\begin{align}
  \ve{\tilde v}\cdot(\ve F-\ve F^{\mathrm{ext}})+\ve{\tilde \omega}\cdot\ve T
  = 
  -6\pi\ve{\tilde v}\cdot\ve v + 8\pi\alpha\ve{\tilde \omega}\cdot(\ve \Omega-\ve \omega) 
  + \ve{\tilde v}\cdot\int_{S_p} \ve\sigma^{E} \cdot \ve {\rd S}
  + \ve{\tilde \omega}\cdot\int_{S_p} \ve r \times (\ve\sigma^{E} \cdot \ve {\rd S}) \nn\\
  - \ve{\tilde v}\cdot\int_V \ma M_{\ve v}\tr \nabla \cdot \ve\sigma^{E}{\rd V}
  - \ve{\tilde \omega}\cdot\int_V \ma M_{\ve \omega}\tr \nabla \cdot \ve\sigma^{E}{\rd V}\,.\eqnlab{rt}
\end{align}
The surface integrals {\lq}at infinity{\rq} do not contribute, and the remaining surface integrals are only over the particle surface. This is well known for the disturbance quantities, say $\ve \sigma^E(\ve u)-\ve \sigma^E(\ve u^\infty)$, because the integrands decay faster than $1/r^2$ \cite{leal1979}. The potentially problematic terms are those from $\ve \sigma^E(\ve u^\infty)$ that are independent of $\ve r$. But $\ma M_{\ve v}$ is an even function of $\ve r$, so that surface integral vanishes by symmetry. On the other hand, $\ma M_{\ve \omega}$ integrated over the sphere is an antisymmetric tensor that vanishes upon contraction with the symmetric stress tensor.

Because $\ve{\tilde v}$ and $\ve{\tilde \omega}$ may be chosen arbitrarily, we have two separate theorems for the force and torque:
\begin{align}
  (\ve F-\ve F^{\mathrm{ext}})
  = 
  -6\pi\ve v
  + \int_{S_p} \ve\sigma^{E} \cdot \ve {\rd S}
  - \int_V \ma M_{\ve v}\tr \nabla \cdot \ve\sigma^{E}{\rd V}\eqnlab{rt_trans}\,.
\end{align}
\begin{align}
  \ve T
  = 
  8\pi\alpha(\ve \Omega-\ve \omega) 
  + \int_{S_p} \ve r \times (\ve\sigma^{E} \cdot \ve {\rd S})
  - \int_V \ma M_{\ve \omega}\tr \nabla \cdot \ve\sigma^{E}{\rd V}\eqnlab{rt_torque}\,.
\end{align}
In this paper we do not consider external torques, and therefore $\ve T=0$ implies
\begin{align}
  \ve\omega = \ve \Omega
  + \frac{1}{8\pi\alpha}\int_{S_p} \ve r \times (\ve\sigma^{E} \cdot \ve {\rd S})
  - \frac{1}{8\pi\alpha}\int_V \ma M_{\ve \omega}\tr \nabla \cdot \ve\sigma^{E}{\rd V}\eqnlab{rt_rot}\,.
\end{align}

The reciprocal theorem may be used to solve either the resistance problem, or the mobility problem. For the resistance problem we take $\ve F^{\mathrm{ext}}=0$ and use \Eqnref{rt_trans}. For the mobility problem we require the total force $\ve F=0$ and find
\begin{align}
  \ve v
  = \frac{1}{6\pi}\ve F^{\mathrm{ext}}
  + \frac{1}{6\pi}\int_{S_p} \ve\sigma^{E} \cdot \ve {\rd S}
  - \frac{1}{6\pi}\int_V \ma M_{\ve v}\tr \nabla \cdot \ve\sigma^{E}{\rd V}\eqnlab{rt_v}\,.
\end{align}
The integrands in Eqs.~(\ref{eqn:rt_trans}-\ref{eqn:rt_v}) are functions of the yet unknown $\ve \sigma^{E}$. In \Secref{results} we evaluate these integrals by calculating $\ve \sigma^E$ perturbatively.

As a final note, \Eqnref{rt_trans} is equivalent to the integral theorem \citet{ho_migration_1976} used to compute the lateral drift of a spherical particle in wall-bounded flow. Their Eq.~(2.22) follows from our \Eqnref{rt_trans} because
\begin{align}
    \int_S \ve {\tilde u} \cdot \ve\sigma^{E} \cdot \ve {\rd S}
  - \int_V \ve{\tilde u}\cdot \nabla \cdot \ve\sigma^{E}{\rd V} =
  \int_V \ve\sigma^{E}:\nabla\ve{\tilde u}{\rd V}\,.
\end{align}
However, we evaluate the two integral contributions in \Eqnref{rt_trans} separately, because they give the contributions from two different physical mechanisms. The surface integral represents the extra polymer stress acting directly on the particle surface (see \Eqnref{v2polymer} and \eqnref{f2polymer} in \Secref{results}). The volume integral represents the indirect effect that the polymer stress modifies the flow field, which in turn modifies the viscous stress on the particle.

\section{Method: T-tensors}\seclab{Ttensors}
In this Section we introduce a basis set of symmetric rank-$n$ Cartesian tensors $T^{nl}_{i_1i_2..i_n}$. Each of these basis tensors is a linear combination of spherical harmonics $Y_l^m$ with a particular value of the angular momentum quantum number $l$, but different modes $m$. Therefore the Cartesian tensors share many useful properties with the spherical harmonics. For example, surface integrals vanish unless $l=0$, tensors with different values of $l$ are orthogonal with respect to integration over the unit sphere, and they have known Fourier transforms.

Because of their direct relation to the spherical harmonics, the $T$-tensors are an alternative basis for Lamb's general solution for Stokes flow \cite{kim1991}. But as explained in detail below, a rank-$n$ $T$-tensor is also closely related to the rank-$n$ polyad $\hat r_{i_1}\hat r_{i_2}..\hat r_{i_n}$ of a unit vector $\ve {\hat r}$. Together with the radial functions $1/r^m$ these polyads are the building blocks of the familiar multipole expansion for Stokes flow, for example the Stokeslet $\delta_{ij}/r + \hat r_i\hat r_j/r$, or the rotlet $\lc_{ijk}\hat r_j/r^2$. Therefore the $T$-tensors stand as a new alternative between Lamb's general solution in spherical coordinates, and the Cartesian multipole expansion. Although any calculation may in principle be performed in any of these representations, we found that the basis described here is suitable for implementation in computer algebra. In particular it enables us to write down particular solutions to inhomogenous Stokes equations in tensorial form, without any explicit coordinate representation, and without explicitly solving differential equations.

In this Section we use index notation to avoid any ambiguity. When appropriate we use the vector notation $\ve T^{nl}$, remembering that $\ve T^{nl}$ is rank $n$ and symmetric in all indices.

\subsection{Definition}
We consider the rank-$n$ polyad $\hat r_{i_1}\hat r_{i_2}..\hat r_{i_n}$ of a unit vector $\ve {\hat r}$. 
Any given polyad is a smooth function defined on the sphere, and may be expanded in the spherical harmonics
\begin{align}
  \hat r_{i_1}\hat r_{i_2}..\hat r_{i_n} &= \sum_{l=0}^{\infty}\sum_{m=-l}^{m=l}c^{nlm}_{i_1i_2..i_n}Y_l^m\,.\eqnlab{Tdefexpansion1}
\end{align}
For our present purposes it is not necessary to calculate the expansion coefficients $c^{nlm}_{i_1i_2..i_n}$, but we may deduce the following important fact.
The left hand side is a polynomial of order $n$, and every term in the sum on the right hand side is a polynomial of order $l$, so we conclude that $c^{nlm}_{i_1i_2..i_n}=0$ if $l>n$. 
Therefore 
\begin{align}
  \hat r_{i_1}\hat r_{i_2}..\hat r_{i_n} &= \sum_{l=0}^{n}\sum_{m=-l}^{m=l}c^{nlm}_{i_1i_2..i_n}Y_l^m\,,\eqnlab{Tdefexpansion2}
\end{align}
We define the tensor $T^{nl}_{i_1i_2..i_n}$ as the inner sum in \Eqnref{Tdefexpansion2}, so that
\begin{align}
T^{nl}_{i_1i_2..i_n} &\equiv \sum_{m=-l}^{m=l}c^{nlm}_{i_1i_2..i_n}Y_l^m\,,\eqnlab{Tdefinition}
\intertext{and therefore by construction}
  \hat r_{i_1}\hat r_{i_2}..\hat r_{i_n} &= \sum_{l=0}^{n}T^{nl}_{i_1i_2..i_n}\,.\eqnlab{polyadexpansion}
\end{align}
We complete the definition by $T^{00} = 1$.

\subsection{Properties of the $T$-tensors}
{\em Symmetry.}
From the expansion \eqnref{polyadexpansion} it is clear that any $T$-tensor is symmetric in all indices. By definition only tensors with $l \leq n$ are non-zero:
\begin{align}
    T^{nl}_{i_1i_2..i_n}&=0\,,\quad l > n\,. 
\end{align}
Further, the polynomial in the left hand side of \Eqnref{polyadexpansion} has parity $(-1)^n$ under inversion of $\ve{\hat r}$, and every term on the right hand side has parity $(-1)^l$. Therefore $\ve T^{nl}$ is non-zero only if both $n$ and $l$ are even, or if both $n$ and $l$ are odd.

{\em Integrals.}
Any two $T$-tensors are orthogonal with respect to integrals over the unit sphere $S$, because the spherical harmonics enjoy this property:
\begin{align}
    \int_S T^{nl_1}_{i_1i_2..i_n}T^{ml_2}_{i_1i_2..i_m}\rd S = 0,\quad l_1 \neq l_2\,.\eqnlab{orthogonality}
\end{align}
It follows that
\begin{align}
    \int_S T^{nl}_{i_1i_2..i_n}\rd S = \int_S T^{nl}_{i_1i_2..i_n}T^{00}\rd S = 0\,,\quad l\neq0\,.\eqnlab{l0integral}
\end{align}

{\em Cartesian rank.}
Taking a trace, i.e. contracting any two indices, of a $T$-tensor lowers its rank $n$ by two:
\begin{align}
    T^{nl}_{i_1i_2..i_n}\delta_{i_{n-1}i_n} &= T^{n-2,l}_{i_1i_2..i_{n-2}}\,.\eqnlab{loweringn}
\end{align}
This follows from \Eqnref{polyadexpansion} and the orthogonality property \eqnref{orthogonality}. 
An important consequence is that $T^{ll}_{i_1i_2..i_l}$ is traceless:
\begin{align}
    T^{ll}_{i_1i_2..i_l}\delta_{i_{l-1}i_l} &= 0\,.\eqnlab{traceless}
\end{align}
Conversely, for $l \leq n-2$, we raise the Cartesian rank $n$ by
\begin{align}
    T^{nl}_{i_1i_2..i_n} &= \frac{2}{n(n+1)-l(l+1)}\big(\delta_{i_1i_2}T^{n-2,l}_{i_3i_4..i_n}+...+\delta_{i_{n-1}i_n}T^{n-2,l}_{i_1i_2..i_{n-3}}\big)\eqnlab{raisingn}
\end{align}
The parenthesis in \Eqnref{raisingn} contains $n(n-1)/2$ terms, one for each unique pairing of the $n$ indices. We have not proven \Eqnref{raisingn} for general values of $n$ and $l$, but it is straightforward to work out the cases $l=n-2$, $l=n-4$, and so on, by taking a trace of \Eqnref{raisingn} and using Eqns.~\eqnref{loweringn} and \eqnref{traceless} repeatedly. We have checked all values of $n$ and $l$ that are used in our calculations in this paper.

{\em Dimensionality.}
The tensor $\ve T^{ll}$ is a rank-$l$ Cartesian tensor. In general it could have $3^l$ unique elements (in three spatial dimensions). In contrast, there are only $2l+1$ spherical harmonics $Y_l^m$ of degree $l$, corresponding to the values $m=-l...\,l$.
But we have shown that $\ve T^{ll}$ is symmetric and traceless. A symmetric tensor of rank $l$ has $(l+1)(l+2)/2$ unique elements, and it has $l(l-1)/2$ unique traces that we require to vanish. These conditions leave exactly $2l+1$ degrees of freedom for a symmetric and traceless rank-$l$ Cartesian tensor. This is the reason we refer to the $T$-tensors as a basis set. The coefficients $c^{llm}_{i_1i_2..i_l}$ in \Eqnref{Tdefinition} are the elements of the \lq rotation matrix{\rq} between the two basis sets $\ve T^{ll}$ and $Y_l^m$. We claim that this transformation is unitary for a certain choice of normalization of the spherical harmonics. In other words, it is in fact a proper rotation. However, we have not proven this for general values of $l$, but we confirmed that it is true up to $l=8$ by brute force calculation of $c^{llm}$ from the definition \Eqnref{Tdefinition}, see \Appref{Tquestions}.

{\em Multiplication.}
The product of two $T$-tensors follows directly from \eqnref{polyadexpansion} as a recurrence relation:
\begin{align}
T^{l_1 l_1}_i T^{l_2 l_2}_j &= 
\sum_{J=0}^{l_1+l_2} T^{l_1+l_2,J}_{\ve i \ve j}
-\sum_{j_1=0}^{l_1-2}\sum_{j_2=0}^{l_2-2} T^{l_1j_1}_{\ve i} T^{l_2j_2}_{\ve j}
-\sum_{j_1=0}^{l_1-2} T^{l_1j_1}_{\ve i} T^{l_2l_2}_{\ve j}
-\sum_{j_2=0}^{l_2-2} T^{l_1l_1}_{\ve i} T^{l_2j_2}_{\ve j}\,.\eqnlab{Tmul}
\end{align}
Here $\ve i$ and $\ve j$ are short-hand for $i_1..i_n$ and $j_1..j_n$. We use \Eqnref{Tmul} in practical calculations, but there is also a largely unexplored connection to quantum angular momentum algebra and Clebsch-Gordan coefficients.
In particular, it can be shown (\Appref{Tquestions}) that
\begin{align}
T^{l_1 l_1}_{\ve i} T^{l_2 l_2}_{\ve j} &= \sum_{J=|l_1-l_2|}^{l_1+l_2}A_{\ve i\ve j\ve k}^{l_1l_1l_2l_2J}T^{JJ}_{\ve k}\,,\eqnlab{TmulCG}
\end{align}
for some coupling tensor $A$ independent of $\ve{\hat r}$.

{\em Relation to polyads.}
We convert any polyadic expression into $T$-tensors by replacing $\hat r_i\to T^{11}_i$, and applying \Eqnref{Tmul} until no products remain. Conversely any $T$-tensor is expressed as a polyadic by recursively using
\begin{align}
    T^{nn}_{i_1i_2..i_n} &= \hat r_{i_1}\hat r_{i_2}..\hat r_{i_n} - \sum_{l=0}^{n-2}T^{nl}_{i_1i_2..i_n}\,,
\end{align}
and \Eqnref{raisingn}. The first few tensors are
\begin{align}
    T^{00} &= 1\,,\quad T^{11}_i = \hat r_i\,\nn\\
    T^{20} &= \frac{1}{3}\delta_{ij}\,,\quad T^{22}_{ij}=\hat r_i\hat r_j-T^{20}_{ij}\nn\\
    T^{31}_{ijk}&=\frac{1}{5}\left(\delta_{ij}\hat r_k+\delta_{ik}\hat r_j+\delta_{jk}\hat r_i\right)\,,\quad T^{33}_{ijk}= \hat r_i\hat r_j\hat r_k - T^{31}_{ijk}\nn\\
    T^{40}_{ijkl} &=\frac{1}{15} \left(\delta _{i l} \delta _{j k}+\delta _{i k} \delta _{j l}+\delta _{i j} \delta _{k l}\right)\nn\\
    T^{42}_{ijkl} &= \frac{1}{7} \left(\hat r_k \hat r_l \delta _{i j}+\hat r_j \hat r_l \delta _{i k}+\hat r_i \hat r_l \delta _{j k}+\hat r_j \hat r_k \delta _{i l}+\hat r_i \hat r_k \delta _{j l}+\hat r_i \hat r_j \delta _{k l}\right)-\frac{2}{21}  \left(\delta _{i l} \delta _{j k}+\delta _{i k} \delta _{j l}+\delta _{i j} \delta _{k l}\right)\nn\\
    T^{44}_{ijkl} &= \hat r_i\hat r_j\hat r_k\hat r_l - T^{42}_{ijkl} - T^{40}_{ijkl}\,.\nn\\
\end{align}

{\em Differentiation.}
In order to calculate using only algebraic manipulations on the $T$-tensors, we must know how the gradient operator $\nabla$, where $\nabla_i = \partial/\partial r_i$, acts on them. We will show that the action of $\nabla$ is to \lq scatter{\rq} a tensor of degree $l$ into a linear combination of tensors with degrees $l-1$ and $l+1$ (given in \Eqnref{difffinal}). We will briefly describe how to compute the coefficients of this linear combination, for any $l$.

Consider the differential operator $T^{nl}_{i}(\nabla)$, defined by taking the polynomial $r^l T^{nl}_i(\ve{\hat r})$ and replacing the components of $\ve  r$ with the partial derivatives $\partial/\partial r_i$. Hobson's theorem on differentiation \cite{hobson_theorem_1892,weniger_spherical_2005} explains that any such differential operator built from a harmonic polynomial acts on radial functions in a particularly simple way. In our case we use his result to find 
\begin{align}
  T^{nl}_{i}(\nabla)\,r^a &= b_a^lr^{a-l}T^{nl}_i(\ve{\hat r})\,, \eqnlab{radialdiff}
\end{align}
with
\begin{align}
  b_a^l &= \prod_{k=0}^{l-1}(a-2k)\,.\eqnlab{radialdiffb}
\end{align}
Therefore the general $J$-th order derivative is
\begin{align}
  T^{NJ}_{i}(\nabla)\, r^m T^{nl}_j(\ve{\hat r})&= \frac{1}{b_{m+l}^l}T^{NJ}_{i}(\nabla)\,T^{nl}_{j}(\nabla)\,r^{m+l}\,.\eqnlab{diffgeneral}
\end{align}
The product $T^{NJ}_{i}(\nabla)\,T^{nl}_{i}(\nabla)$ is given by \Eqnref{TmulCG}, and the general formula follows from \Eqnref{radialdiff}. In this paper we only consider first order derivatives which correspond to $J=1$, because $\partial/\partial r_i = T_i^{11}(\nabla)$. For $J=1$ the general formula \eqnref{diffgeneral} and \Eqnref{TmulCG} give
\begin{align}
  \frac{\partial}{\partial r_i} r^m T^{nl}_j(\ve{\hat r})&= 
  \frac{1}{b_{m+l}^l}\nabla^2 A_{ijk}^{11nl,l-1}T^{l-1,l-1}_k(\nabla)\,r^{m+l} +
  \frac{1}{b_{m+l}^l} A_{ijk}^{11nl,l+1}T^{l+1,l+1}_k(\nabla)\,r^{m+l}\,,
\end{align}
which becomes, using \Eqnref{radialdiff} and \Eqnref{radialdiffb} ,
\begin{align}
   = (m+l+1)A_{ijk}^{11nl,l-1}T^{l-1,l-1}_k(\ve{\hat r})r^{m-1} +
   (m-l)A_{ijk}^{11nl,l+1}T^{l+1,l+1}_k(\ve{\hat r})r^{m-1}\,.\eqnlab{difffinal}
\end{align}
To evaluate \Eqnref{difffinal} in our computer program we compute the product $r^{m-1}\ve T^{11}\ve T^{nl}\equiv\alpha r^{m-1} \ve T^{l-1,l-1} + \beta r^{m-1}\ve T^{l+1,l+1}$ using \eqnref{Tmul}, and replace the coefficients by $\alpha \rightarrow(m+l+1)\alpha$, and $\beta\rightarrow(m-l)\beta$.

{\em Fourier transform.}
We use a symmetric convention for the Fourier transform:
\begin{align}
\fourier{f}(\ve k) &\equiv \frac{1}{(2\pi)^{3/2}}\int_{\mathbb R^3}\rd^3\ve r e^{-i \ve r\cdot \ve k}f(\ve r)\,,\\
\invfourier{f}(\ve r) &\equiv \frac{1}{(2\pi)^{3/2}}\int_{\mathbb R^3}\rd^3\ve k e^{i \ve r\cdot \ve k}f(\ve k)\,.
\end{align}
The Fourier transform of $r^m T^{nl}_i(\ve{\hat r})$ follows directly from that of the functions $r^m Y_l^\mu(\theta,\varphi)$ given in Ref.~\cite{samko_fourier_1978}. For all values of $m$ and $l$ that appear in the present calculation
\begin{align}
    \fourier r^m T^{nl}(\ve{\hat r}) &= \frac{\Psi_{ml}}{k^{m+3}}T^{nl}(\ve{\hat k})\,,\quad m\neq l+2j \textrm{ and } m\neq -(l+3)-2j\,,\quad j=0,1,... \eqnlab{fourier}\\
    \Psi_{ml}&= (-i)^l 2^{m+3/2}\frac{\Gamma(\frac{m+l+3}{2})}{\Gamma(\frac{l-m}{2})}
\end{align}
When $m=l+2j$, $r^m T^{nl}(\ve {\hat r})$ is a polynomial in the components of $\ve r$, and its Fourier transform is the Dirac delta function and its derivatives. The case $m= -(l+3)-2j$ is more complicated, involving logarithms \cite{samko_fourier_1978}. Neither of these cases arise in this paper.

\subsection{Particular solution for the inhomogenous Stokes equation}\seclab{Tparticular}
Consider the inhomogenous Stokes problem
\begin{align}
  -\partial_i p + \nabla^2 u_i &= f_i\,,\quad \partial_i u_i = 0, \eqnlab{inhomogenous1}
\end{align}
where we assume that $f_i$ is a linear combination of $T$-tensors.
The Fourier transform of \Eqnref{inhomogenous1} is
\begin{align}
    -ik_i\fourier p-k^2\fourier{u_i} &= \fourier  f_i\,,\quad k_i \fourier u_i=0\,,
\end{align}
where $k=|\ve k|$. This algebraic equation is solved by
\begin{align}
    \fourier p &= -\frac{ k_j\fourier f_j}{ik^2}\,,\eqnlab{pufourierp}\\
    \fourier u_i &= -\frac{1}{k^2}(\delta_{ij} - \hat k_i\hat k_j)\fourier f_j\,,\eqnlab{pufourieru}
\end{align}
where $\hat k_i\equiv k_i/k$ is a unit vector. In terms of $T$-tensors, the Fourier space Green's function is
\begin{align}
    -\frac{1}{k^2}(\delta_{ij} - \hat k_i\hat k_j) &= \frac{1}{k^2}\left(T^{22}_{ij}(\ve{\hat k})-\frac{2}{3}\delta_{ij}\right)\,.\eqnlab{greensk}
\end{align}
The procedure to find the solution $u_i$ is therefore
\begin{enumerate}
    \item compute the Fourier transform of $f_i$ using \Eqnref{fourier},
    \item multiply with the Greens function \eqnref{greensk} using \Eqnref{Tmul},
    \item Inverse Fourier transform the product again using \Eqnref{fourier}.
\end{enumerate}
In this paper we never need an explicit expression for the pressure $p$, but if needed it is computed in the analogous way from \Eqnref{pufourierp}.

\section{Results}\seclab{results}
In this Section we give the solutions to both the mobility problem (\Secref{mobility}) and the resistance problem (\Secref{resistance}) for a freely rotating spherical particle in an unbounded viscoelastic shear flow, with no restriction on the direction of $\ve v$ or $\ve F^{\mathrm{ext}}$ relative to the shear. 

\subsection{The mobility problem}\seclab{mobility}
Here we consider a particle moving under the effect of an external force. The particle velocity $\ve v$ is a function of $\De$ and $\Wi$ to be determined, and to that end we require that the total force $\ve F=0$. We proceed with a regular perturbation expansion in $\De$:
\begin{align}
  \ve u &= \ve u^{(0)}+\De\, \ve u^{(1)}+\De^2\, \ve u^{(2)}+...\nn\\
  p &= p^{(0)}+\De\, p^{(1)}+\De^2\, p^{(2)}+...\nn\\
  \ve \omega &= \ve \omega^{(0)}+\De\, \ve \omega^{(1)}+\De^2\, \ve \omega^{(2)}+...\nn\\
  \ve v &= \ve v^{(0)}+\De\, \ve v^{(1)}+\De^2\, \ve v^{(2)}+...\nn\\
  \ve \Pi &= \ve \Pi^{(0)}+\De\, \ve \Pi^{(1)}+\De^2\, \ve \Pi^{(2)}+...\nn
\end{align}
At each order $\ve \Pi^{(k)}$ is given by an algebraic equation, and $\ve u^{(k)}$ by an inhomogenous Stokes equation (except for the lowest order, which is homogenous).
To lowest order $\De^0$ we have from \Eqnref{oldBdef} 
\begin{align}
    \ve \Pi^{(0)} &= \nabla \ve u^{(0)}+(\nabla \ve u^{(0)})\tr\,,
\end{align}
and therefore $\ve\sigma^{E(0)}=0$, and therefore from Eqs.~\eqnref{rt_v} and \eqnref{rt_rot} we have
\begin{align}
\ve v^{(0)} &= \frac{1}{6\pi}\ve F^{\mathrm{ext}}\,,\nn\\
\ve\omega^{(0)} &= \ve \Omega
\end{align}
To order $\De^0$ the flow satisfies
\begin{align}
  &-\nabla p^{(0)} + \nabla^2 \ve u^{(0)} = 0\,,\\
  \ve u^{(0)} &= \alpha \ve \omega^{(0)} \times \ve r,\quad\ve r \in S\nn\\
  \ve u^{(0)} &\to  \alpha\ve u^\infty-\ve v^{(0)}\,,\quad|\ve r|\to \infty\,. 
\end{align}
At order $\De^1$ \Eqnref{oldBdef} gives
\begin{align}
\label{eq:eqPi1}
\ve \Pi^{(1)} &= -\left[(\ve u^{(0)}\cdot\nabla)\ve \Pi^{(0)} - (\nabla \ve u^{(0)})\ve \Pi^{(0)} - \ve \Pi^{(0)} (\nabla \ve u^{(0)})\tr\right] + \nabla \ve u^{(1)} + (\nabla \ve u^{(1)})\tr\,,
\end{align}
where $\ve u^{(0)}$ is known, but $\ve u^{(1)}$ is still unknown.
Consequently
\begin{align}
\ve \sigma^{E(1)} &= -\left[(\ve u^{(0)}\cdot\nabla)\ve \Pi^{(0)} - (\nabla \ve u^{(0)})\ve \Pi^{(0)} - \ve \Pi^{(0)} (\nabla \ve u^{(0)})\tr\right] \,.\eqnlab{sigmae1}
\end{align}
The reciprocal theorem \eqnref{rt_v} gives
\begin{align}
  \ve v^{(1)}
  &= \frac{1}{6\pi}\int_{S_p} \ve\sigma^{E(1)} \cdot \ve {\rd S}
  - \frac{1}{6\pi}\int_V \ma M_{\ve v}\tr \nabla \cdot \ve\sigma^{E(1)}{\rd V} =-\frac{\alpha\mu_r}{6\pi} \ve \Omega\times \ve F^{\mathrm{ext}}\eqnlab{mobilityvorder1}\\
  \ve\omega^{(1)} &= \frac{1}{8\pi\alpha}\int_{S_p} \ve r \times (\ve\sigma^{E(1)} \cdot \ve {\rd S})
  - \frac{1}{8\pi\alpha}\int_V \ma M_{\ve \omega}\tr \nabla \cdot \ve\sigma^{E(1)}{\rd V} = 0\,.\eqnlab{mobilityomegaorder1}
\end{align}
At this order the shear flow and the disturbance from the external forcing interact to create a lateral drift perpendicular to both the direction of forcing and the vorticity. The drift arises from both the extra stress on the particle surface, and the viscous stress induced by the viscoelastic medium, in proportion $2:3$.
The lateral drift \Eqnref{mobilityvorder1} was first calculated by \citet{brunn1977}. Our \Eqnref{mobilityvorder1} agrees with his when accounting for the erratum \cite{brunn1977a} and letting his $\kappa_0^{11}=-2\kappa_0^{22}$, which corresponds to the second-order fluid limit of the Oldroyd-B model. The mobility derived in Ref.~\citenum{vishnampet2012} is different from \Eqnref{mobilityvorder1}. In particular they report a contribution proportional to $\ma S \ve F^{\mathrm{ext}}$.
The $O(\Wi)$ contribution to the angular velocity vanishes, in agreement with all the previous results \cite{brunn1977,brunn1977a,vishnampet2012}.

With $\ve \sigma^{E(1)}$, $\ve v^{(1)}$, and $\ve \omega^{(1)}$ given by Eqs.~(\ref{eqn:sigmae1}-\ref{eqn:mobilityomegaorder1}) we can write down the inhomogenous Stokes problem for $\ve u^{(1)}$:
\begin{align}
  -\nabla p^{(1)} + \nabla^2 \ve u^{(1)} &= -\nabla \cdot \ve \sigma^{E(1)}\,,\eqnlab{order1floweq1}
\end{align}
subject to
\begin{align}
  \ve u^{(1)} &= \alpha \ve \omega^{(1)} \times \ve r,\quad\ve r \in S\nn\\
  \ve u^{(1)} &\to  -\ve v^{(1)}\,,\quad|\ve r|\to \infty\,. \eqnlab{order1flowbc1}
 \end{align}
First we compute a particular solution $\ve u^{(1)p}(\ve r)$ as explained in \Secref{Tparticular}. The flow field $\ve u^{(1)p}$ satisfies the inhomogenous \Eqnref{order1floweq1}, but not the boundary conditions \Eqnref{order1flowbc1}. We next solve for a Stokes flow $\ve u^{(1)h}$ that satisfies the homogenous equation
\begin{align}
  &-\nabla p^{(1)h} + \nabla^2 \ve u^{(1)h} = 0\,,
\end{align}
and the boundary conditions
\begin{align}
  \ve u^{(1)h} &= \ve \omega^{(1)} - \ve u^{(1)p} \times \ve r,\quad\ve r \in S\nn\\
  \ve u^{(1)h} &\to  -\ve u^{(1)p}\,,\quad|\ve r|\to \infty\,. 
 \end{align}
By construction $\ve u^{(1)}=\ve u^{(1)h}+\ve u^{(1)p}$.

At order $\De^2$ we have from \Eqnref{oldBdef}
\begin{align}
\ve \Pi^{(2)} &= 
\ve \sigma^{E(2)}+ \nabla \ve u^{(2)} + (\nabla \ve u^{(2)})\tr\,,
\end{align}
with
\begin{align}
\ve \sigma^{E(2)} &=-\bigg[
(\ve u^{(0)}\cdot\nabla)\ve \Pi^{(1)} - (\nabla \ve u^{(0)})\ve \Pi^{(1)} - \ve \Pi^{(1)} (\nabla \ve u^{(0)})\tr +
(\ve u^{(1)}\cdot\nabla)\ve \Pi^{(0)} - (\nabla \ve u^{(1)})\ve \Pi^{(0)} - \ve \Pi^{(0)} (\nabla \ve u^{(1)})\tr
\bigg] \,,
\end{align}
where $\ve u^{(0)}$, $\ve u^{(1)}$,  $\ve \Pi^{(0)}$ and $\ve \Pi^{(1)}$ are all known. The reciprocal theorem at order $\De^2$ gives
\begin{align}
  \ve v^{(2)}
  &= \frac{1}{6\pi}\int_{S_p} \ve\sigma^{E(2)} \cdot \ve {\rd S}
  - \frac{1}{6\pi}\int_V \ma M_{\ve v}\tr \nabla \cdot \ve\sigma^{E(2)}{\rd V} \eqnlab{v2rt}\\
  \ve\omega^{(2)} &= \frac{1}{8\pi\alpha}\int_{S_p} \ve r \times (\ve\sigma^{E(2)} \cdot \ve {\rd S})
  - \frac{1}{8\pi\alpha}\int_V \ma M_{\ve \omega}\tr \nabla \cdot \ve\sigma^{E(2)}{\rd V} \eqnlab{omega2rt}
\end{align}
We first consider the angular velocity $\ve \omega^{(2)}$. The angular velocity due to the surface integral in \Eqnref{omega2rt} vanishes, so the induced viscous stress alone explains the rotation rate at this order.
The full result for the angular velocity, to second order in $\De$ and $\Wi$, takes the form
\begin{align}
    \ve \omega &= \ve\Omega - \Wi^2\,\frac{\mu_r}{2} \ve\Omega + \De^2\,  \frac{5\mu_r (910 \mu_r-1941)}{96096} \frac{1}{(6\pi)^2}\ve F^\mathrm{ext} \times \ma S \ve F^\mathrm{ext}\,.\eqnlab{mobilityomega}
\end{align}
The numerically largest contribution is the $O(\Wi^2)$ slowdown of the rotation around vorticity. This contribution agrees with an earlier analytical result \cite{housiadas2011a} that also explains numerical simulations \cite{davino2008}. The $O(\De^2)$ contribution shows a coupling between the external force and rotation rate, through the strain. It is numerically small, but may be important because it describes a rotation around another axis than $\ve \Omega$.

For the particle velocity $\ve v^{(2)}$ the surface integral of the extra stress on the particle evaluates to
\begin{align}
   \int_{S_p} \ve\sigma^{E(2)} \cdot \ve {\rd S}&=\alpha^2\mu_r\bigg(
    \frac{2}{3}(\mu_r-1) \ve \Omega\times\ve \Omega\times\ve F^\mathrm{ext}
    -\frac{1}{3}\,\ve \Omega \times \ma S \ve F^\mathrm{ext}\eqnlab{v2polymer}
    \bigg)\,.
\end{align}
We see that this contribution may affect the velocity along $\ve F^\mathrm{ext}$, but in particular it gives another lateral drift in a direction perpendicular to the $O(\Wi)$ lateral drift calculated above.

Next we evaluate the volume integral in \eqnref{v2rt}, and this gives the final result for $\ve v$ to second order in $\De$ and $\Wi$: 
\begin{align}
    6\pi \ve v &= \ve F^\mathrm{ext} - \Wi\,  \mu_r\ve \Omega\times \ve F^\mathrm{ext} +
    \mu_r\bigg(
    \De^2\, \frac{143 \mu_r+258}{25025}\frac{|\ve F^\mathrm{ext}|^2}{(6\pi)^2}
    + \Wi^2\,  \frac{5  (237005 \mu_r-291618)}{378378}|\ve \Omega|^2
    \bigg)\ve F^\mathrm{ext}\nn\\
    &\qquad+\Wi^2\mu_r\bigg((\mu_r-1) \ve \Omega\times\ve \Omega\times\ve F^\mathrm{ext}
    + \frac{3}{2}\,\ve \Omega \times \ma S \ve F^\mathrm{ext}
    + \frac{183339-286735 \mu_r}{126126} \ma S \ma S \ve F^\mathrm{ext}\bigg)
    \eqnlab{mobilityv}
\end{align}
The induced viscous stress, given by the volume integral, contributes to the same terms as the extra stress on the surface shown in \Eqnref{v2polymer}. In addition there is yet another velocity proportional to $\ma S\ma S\ve F^\mathrm{ext}$, and a component in the direction of $\ve F^\mathrm{ext}$.

The second order contribution to the velocity directly proportional to $\ve F^\mathrm{ext}$ consists of one term proportional to $\De^2$, and one proportional to $\Wi^2$. The velocity along $\ve F^\mathrm{ext}$ increases as $\De$ increases, but the numerical prefactor is small. The important result is that the velocity decreases with increasing shear rate, as observed in experiment \cite{van_den_brule_effects_1993}. We discuss this effect for an inclined shear flow in \Secref{discussion}. In the next Section we solve the resistance problem that can be directly compared with earlier calculations for the cross-shear flow.

\subsection{The resistance problem}\seclab{resistance}
The calculation for this problem is very similar to that of the mobility problem, so we omit most details. Here we consider a freely rotating sphere moving at velocity $\ve v$ through a shear flow, and calculate the resulting hydrodynamic force $\ve F$ and angular velocity $\ve \omega$. The crucial differences to the mobility problem are that $\ve F^{\mathrm{ext}}=0$, and $\ve v$ is a constant, independent of $\De$.

The zeroth order problem is the corresponding Stokes problem, which determines $\ve u^{(0)}$ and $\ve \sigma^{E(1)}$. The reciprocal theorem \eqnref{rt_rot} gives
\begin{align}
\ve\omega^{(0)} &= \ve \Omega\,, \nn\\
  \ve\omega^{(1)} &= \frac{1}{8\pi\alpha}\int_{S_p} \ve r \times (\ve\sigma^{E(1)} \cdot \ve {\rd S})
  - \frac{1}{8\pi\alpha}\int_V \ma M_{\ve \omega}\tr \nabla \cdot \ve\sigma^{E(1)}{\rd V} = 0\,.
\end{align}
The first order flow problem has a different boundary condition to that of the mobility problem, because $\ve v$ is independent of $\De$. The first order equations are
\begin{align}
  -\nabla p^{(1)} + \nabla^2 \ve u^{(1)} &= -\nabla \cdot \ve \sigma^{E(1)}\,,\eqnlab{order1floweqres}
\end{align}
subject to
\begin{align}
  \ve u^{(1)} &= \alpha \ve \omega^{(1)} \times \ve r,\quad\ve r \in S\nn\\
  \ve u^{(1)} &\to  0\,,\quad|\ve r|\to \infty\,. \eqnlab{order1flowbcres}
 \end{align}
The reciprocal theorem \eqnref{rt_trans} gives
\begin{align}
  \ve F^{(0)}
  &= -6\pi\ve v\,, \nn\\
  \ve F^{(1)}
  &= \int_{S_p} \ve\sigma^{E(1)} \cdot \ve {\rd S}
  - \int_V \ma M_{\ve v}\tr \nabla \cdot \ve\sigma^{E(1)}{\rd V}\,, \nn\\
  \ve F^{(2)}
  &= \int_{S_p} \ve\sigma^{E(2)} \cdot \ve {\rd S}
  - \int_V \ma M_{\ve v}\tr \nabla \cdot \ve\sigma^{E(2)}{\rd V} \,,\nn\\
  \ve\omega^{(2)} &= \frac{1}{8\pi\alpha}\int_{S_p} \ve r \times (\ve\sigma^{E(2)} \cdot \ve {\rd S})
  - \frac{1}{8\pi\alpha}\int_V \ma M_{\ve \omega}\tr \nabla \cdot \ve\sigma^{E(2)}{\rd V}\,. \eqnlab{rtF2}
\end{align}
The contributions from the surface and volume integrals are similar to those of the mobility problem. Specifically, 
\begin{align}
    \frac{1}{6\pi}\int_{S_p} \ve\sigma^{E(2)} \cdot \ve {\rd S} &= 
    \alpha^2\,\mu_r\bigg(
    -\frac{2}{3}\ve\Omega\times\ve\Omega\times\ve v
    -\frac{\mu_r}{3}\ve\Omega\times\ma S\ve v
    \bigg)\,.\eqnlab{f2polymer}
\end{align}
After evaluating the volume integrals in \Eqnref{rtF2}, we find for the resistance problem with velocity $\ve v$
\begin{align}
    \frac{\ve F}{6\pi} &= -\ve v-\Wi\,\mu_r \ve\Omega \times \ve v
 +\mu_r\bigg(\De^2\,\frac{143 \mu_r+258}{25025}|\ve v|^2+\Wi^2\,\frac{5 (237005 \mu_r-291618)}{378378}|\ve \Omega|^2\bigg)\ve v \nn\\
    &\quad+\Wi^2\,\mu_r\bigg(
    -\ve\Omega\times\ve\Omega\times\ve v
    +\frac{3}{2}\ve\Omega\times\ma S\ve v
    +\frac{183339-286735 \mu_r}{126126}\ma S\ma S\ve v
    \bigg)\eqnlab{resistanceF}\\
    \ve \omega &= \ve \Omega - \Wi^2\,\frac{\mu_r}{2}\ve \Omega + \De^2\,\frac{5\mu_r (910 \mu_r-1941)}{96096}\ve v\times\ma S\ve v
\end{align}
As expected, the expression for the resistance force \eqnref{resistanceF} is similar to the expression for the mobility velocity \eqnref{mobilityv} with $\ve F^\mathrm{ext}$ replaced by $-6\pi\ve v$. However, they differ in a term $\Wi^2 \mu_r^2\,\ve\Omega\times\ve\Omega\times\ve v$. This difference arises because of the lateral force at $O(\Wi)$ for the following reason. In the mobility problem the particle is allowed to relax this lateral hydrodynamic force by drifting sideways. But in the resistance problem we essentially force the fluid, through the boundary conditions, with the lateral force required to keep the particle moving with the prescribed velocity $\ve v$. This forcing, or lack thereof, at $O(\Wi)$ is what gives the differing term at $O(\Wi^2)$. Upon substitution of the mobility velocity \Eqnref{mobilityv} into the expression for the resistance force \Eqnref{resistanceF} we find $\ve F = -\ve F^{\mathrm{ext}}$ to second order in $\De^2$ and $\Wi^2$, as advertised in the Introduction.

The drag term proportional to $\De^2$ is the drag in absence of shear. This term was first calculated by \citet{leslie1961} in the context of sedimentation in a quiescent fluid. Our coefficient matches theirs when $\mu_r=1$, and their $\epsilon=\beta=0$.

For the cross-shear flow, $\ve v$ is parallel to $\ve \Omega$, so that $\ve \Omega \times \ve v=\ma S\ve v=0$. For this case only the first and third terms on the right-hand side of \Eqnref{resistanceF} remain. This expression agrees with the analytical result of \citet{housiadas_rheological_2014}, and therefore with \citet{padhy_simulations_2013} as shown in their Fig.~7.

\section{Summary and Discussion}\seclab{discussion}
\begin{figure}
    \begin{overpic}[width=4.5cm]{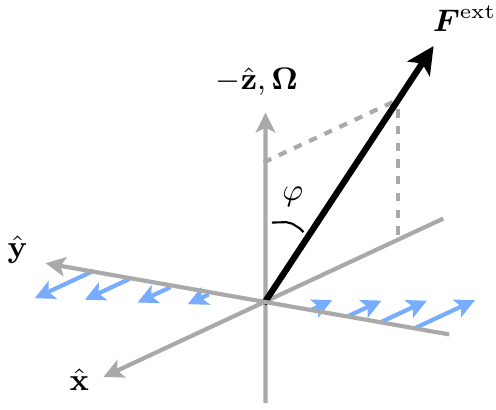}
    \end{overpic}
    \caption{\figlab{inclinedshear} The inclined shear flow geometry discussed in \Secref{discussion}. In this example the external force lies in the plane spanned by the vorticity axis and the flow direction. This situation corresponds to settling between two far-apart shearing walls, parallel to the walls, but where the shearing is at an angle to gravity. We denote by $\varphi$ the angle between $\ve F^\mathrm{ext}$ and vorticity.}
\end{figure}

\begin{figure}  
    \includegraphics{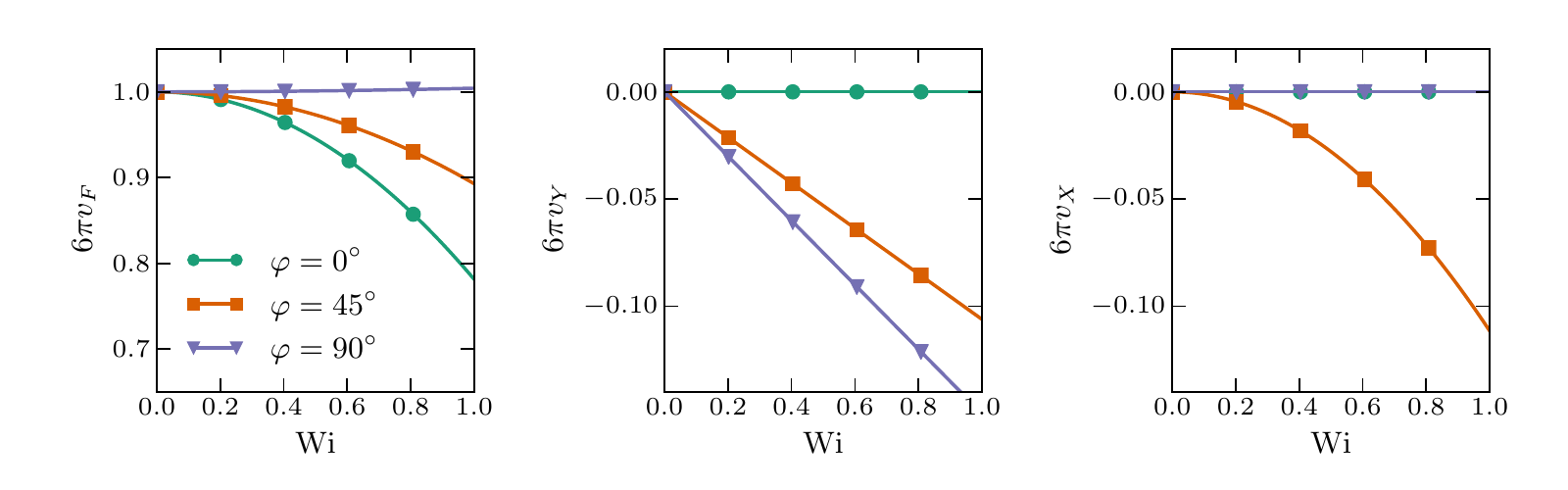}
    \caption{\figlab{v_tilty} Shear-dependent velocity of a sphere forced by an external force in the flow-vorticity plane, for different angle of attack $\varphi$ between the forcing $\ve F^\mathrm{ext}$ and the vorticity axis. The cross-shear flow corresponds to $\varphi=0$, whereas the force is along the flow direction when $\varphi=90^\circ$. (a) Velocity along $\ve F^\mathrm{ext}$. (b) Lateral drift in the shear direction $\ve{\hat y}$.  (c) Lateral drift perpendicular to the shear direction. Parameters: $\mu_r = 0.3$, $\De=0.1$.}
\end{figure}    
   
\begin{figure}
\includegraphics{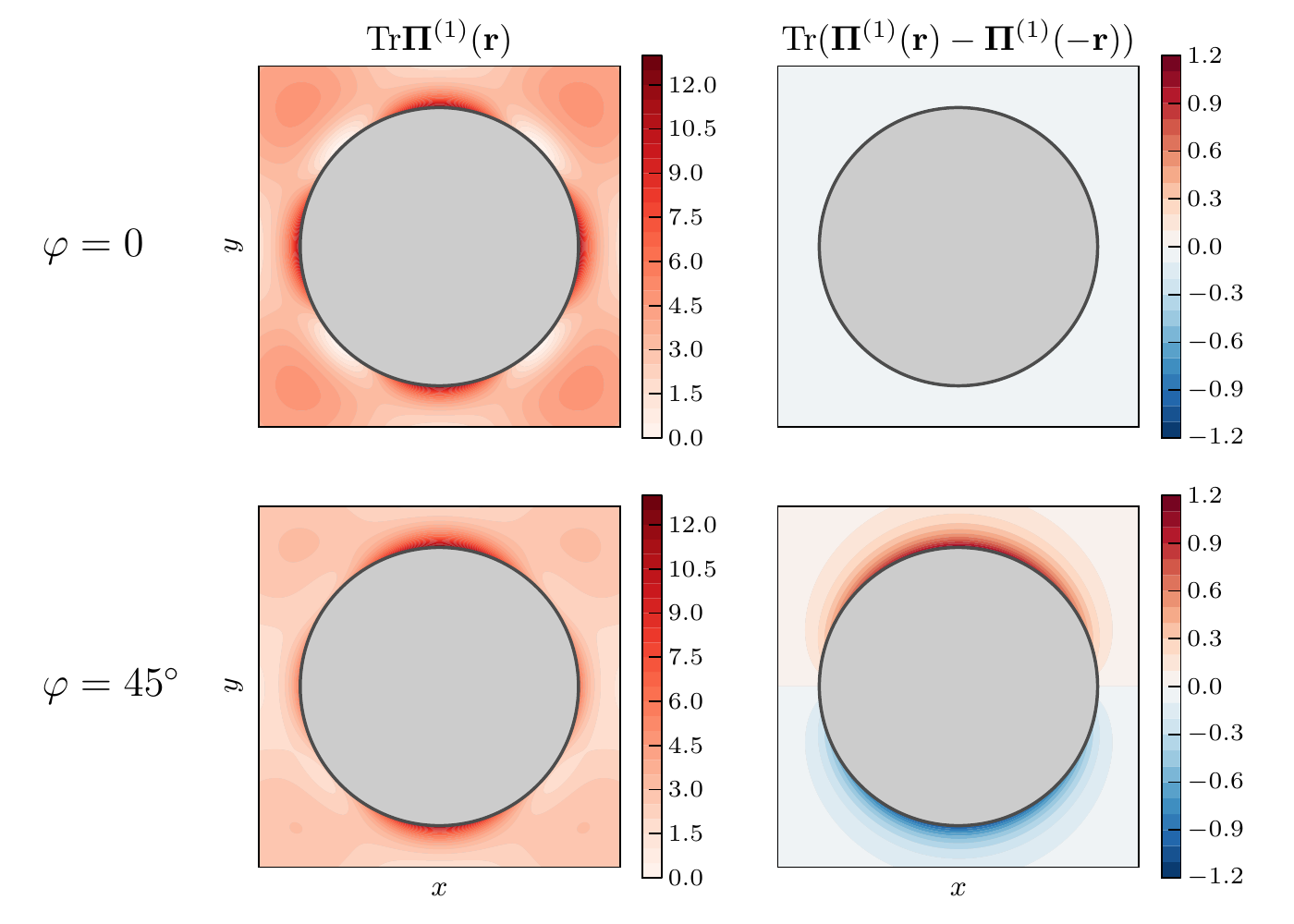}
\caption{\figlab{stress_tilty}
The trace of the first order viscoelastic stress $\ve \Pi^{(1)}$ around a particle driven by an external force through a shear flow. All panels show a center cross-section of the particle, viewed along the direction of external forcing. In these panels $y$ indicates the shear direction, and $x$ indicates the direction perpendicular to both $\ve F^\mathrm{ext}$ and $\ve{\hat y}$, see \Figref{inclinedshear}. Left column shows full stress field, right column shows its asymmetric part under inversion of $\ve r$. Top row shows stress field when the external forcing is aligned with vorticity, bottom row shows same when the external forcing is at angle $\varphi = 45^\circ$ to vorticity.
Parameters: $\alpha=1$ (implying $\Wi=\De$). The trace of $\ve \Pi^{(1)}$ is independent of $\mu_r$, which follows from Eq.~(\ref{eq:eqPi1}) because the flow field $\ve u$ is incompressible.
}
\end{figure}

We have derived analytical results for the linear and angular velocities of a particle driven through a viscoelastic shear flow by an external force, valid to second order in $\De$ and $\Wi$, given in Eqs.~\eqnref{mobilityv} and \eqnref{mobilityomega}.

We found three qualitatively different corrections to the predicted velocity in a Newtonian fluid. First, at $O(\Wi)$ there is a drift proportional to $\ve \Omega \times \ve F^\mathrm{ext}$, that is, perpendicular to the forcing and vorticity. Second, the resulting velocity along the forcing is modified at $O(\De^2)$ and $O(\Wi^2)$. The numerical prefactor of the $\De^2$-contribution is small, so in practice only the $O(\Wi^2)$ effect is important. These terms correspond to the effect of the imposed shear flow. Third, at $O(\Wi^2)$ there is yet another lateral drift, perpendicular to the first one. Even for $\Wi=0.5$ this second drift may be as strong as the $O(\Wi)$-drift, but it points in another direction. The relative importance of these three effects depends strongly on the direction of external forcing relative to the orientation of the shear flow.

There are two mechanisms that contribute to these corrections. First, the extra stress acts directly on the particle, giving a force and a torque. Secondly, the extra stress acts on the fluid, which modifies the flow and indirectly gives a force and a torque via the viscous and pressure terms. The lateral drift at $O(\Wi)$ is a combination of these two mechanisms [\Eqnref{mobilityomegaorder1}]. The decreased velocity of a sphere sedimenting in a cross-shear flow, however, is due to the indirect increase of viscous stress, and the direct contribution from the extra stress vanishes [\Eqnref{v2polymer}]. This observation is in qualitative agreement with the observations of numerical simulations \cite{padhy_simulations_2013}. When the forcing is at an angle to the vorticity vector, the correction is typically a combination of the two mechanisms.

The angular velocity around the vorticity slows down at $O(\Wi^2)$, in agreement with earlier results \cite{housiadas2011a,davino2008}. But at $O(\De^2)$we also find a coupling between the strain and translation that induces a rotation around the axis $\ve F^{\mathrm {ext}}\times\ma S\ve F^{\mathrm {ext}}$. The prefactor is quite small, but the effect could be important because it describes rotation around another axis than $\ve \Omega$.

{\em Settling in inclined shear flow.} \Eqnref{mobilityv} is valid for any orientation of the forcing relative to the shear flow, described for instance by two angles relative to the vorticity and flow directions. In the remainder of this discussion we focus on the concrete example of a particle settling under gravity, $\ve F^\mathrm{ext}=m\ve g$, with the particular set of orientations so that gravity lies in the plane spanned by the vorticity axis and the flow direction, see \Figref{inclinedshear}. This situation corresponds to settling between two far-apart shearing walls, parallel to the walls, but where the shearing is at an angle to gravity. We denote by $\varphi$ the angle between $\ve g$ and the vorticity, see \Figref{inclinedshear}. 

We show the resulting settling velocity as a function of $\Wi$ in \Figref{v_tilty}, for $\varphi=0$, $45^\circ$, and $90^\circ$. When $\varphi=0$ we recover the cross-shear result, and the lateral drift vanishes. As the angle of inclination increases the settling velocity increases, diminishing the shear-induced drag increase described for $\varphi=0$ in Refs.~\cite{van_den_brule_effects_1993,housiadas_rheological_2014,padhy_simulations_2013}. When gravity acts along the flow direction, $\varphi=90^\circ$, the settling velocity is almost the same as that given by Stokes law, only slightly higher. The direction of the $O(\Wi)$ lateral drift is along the $-\ve{\hat y}$ direction, see \Figref{inclinedshear}.  For finite $\varphi$ and small $\Wi$ this drift is the dominant feature of the particle velocity. But even for larger $\Wi$ the magnitude of this drift is comparable to the reduction in settling velocity when $\varphi=45^\circ$. The additional $O(\Wi^2)$ drift is in the third independent direction, given by $ \ve F^\mathrm{ext}\times\ve{\hat y}$ (\Figref{inclinedshear}). For $\Wi\approx0.5$ it is comparable in magnitude to both the reduction in settling velocity and the $O(\Wi)$ drift.

The direction of the $O(\Wi)$ lateral drift can be understood by considering how the elastic fluid is stretched in the vicinity of the sphere. In \Figref{stress_tilty} we show the trace of the first order elastic stress tensor, $\Tr\ve\Pi^{(1)}(\ve r)$, around the sphere. This trace indicates how strongly the dumbells are stretched by the lowest-order Newtonian flow. The Figure shows a center cross-section of the particle, viewed along the direction of external forcing. In the cross-shear flow, $\varphi=0$, the stretching is a complicated function of the spatial variables, but perfectly symmetric around the sphere (top row in \Figref{stress_tilty}). Therefore there is no net force on the particle. But as the external forcing is tilted, the particle is forced to move along the flow direction. Now the particle surface moves opposite to the undisturbed flow on one side, and along the undisturbed flow on the other. This asymmetry results in different stretching of the dumbbells on the two sides, as illustrated for $\varphi=45^\circ$ in the second row of \Figref{stress_tilty}. This stress contributes to the particle drift both directly, by forcing the particle surface, and indirectly by forcing the suspending fluid and thereby inducing additional viscous drag.

{\em Method.} In this paper we also introduced the tensors $T^{nl}_{i_1i_2..i_n}$ (\Secref{Ttensors}). These tensors are a basis suitable for symbolic calculations of tensorial quantities in spherical geometry. In particular they allow us to write down solutions to inhomogenous Stokes equations in tensorial form, without any explicit coordinate representation, and without explicitly solving differential equations. The calculation in this paper demonstrates the power of our method for treating tensorial equations such as the coupled rank-$2$ constitutive \Eqnref{oldBdef} and the rank-$1$ flow \Eqnref{floweq1}. Nevertheless, there are many open questions regarding the algebraic properties of the $T$-tensors. Most importantly, we have shown that the product $\ve T^{l_1l_1}\ve T^{l_2l_2}$ is given by a linear combination of $\ve T^{JJ}$ with $|l_1-l_2|\leq J\leq l_1+l_2$, analogous to the product of two spherical harmonics (see \Appref{Tquestions}). But further work must be done to determine the properties of the coefficients in this linear combination, to determine the general expression for differentiation, and to prove the general case of \Eqnref{raisingn}. 

Our tensor formalism can also be extended to other geometries. Nearly spherical geometry can be treated by perturbation theory. Other geometries can be treated by the method of images \cite{blake_note_1971,chwang_hydromechanics_1975}. For example, the flow around a spheroid in unbounded flow is given by a finite distribution of multipoles \cite{chwang_hydromechanics_1975}. However, the radial functions are no longer simply $r^m$, but integrals $I_m^n=\int_{-c}^c \xi^n/|\ve r-\xi \ve n|^m\,\rd \xi$, where $\ve n$ is the direction of the spheroid and $c$ is a shape-dependent constant \cite{chwang_hydromechanics_1975,einarsson_rotation_2015}. Their algebraic properties must be worked out in order to use the formulae in \Secref{Ttensors} e.g. for the Fourier transform. For wall interactions, or other many-center problems, it is possible to derive a translation theorem that expresses $|\ve r-\ve r'|^m \ve T^{nl}(\ve r-\ve r')$ as an infinite series of $|\ve r|^m \ve T^{nl}(\ve r)$ and $|\ve r'|^m \ve T^{nl}(\ve r')$ \cite{weniger_spherical_2005}, which restores the linearity of the problem.

{\em Acknowledgements}. This work was supported by Vetenskapsr\aa{}det, and by the grant {\em Bottlenecks for particle growth in turbulent aerosols} from the Knut and Alice Wallenberg Foundation, grant number 2014.0048.



%


\appendix

\section{Multiplication of $T$-tensors}\applab{Tquestions}
We use following normalization of the spherical harmonics,
  \begin{align}
    Y_l^m(\theta, \varphi)&=\sqrt{\sqrt{\pi}\frac{2^{-l} l! \Gamma (l-m+1)}{\Gamma \left(l+\frac{1}{2}\right) \Gamma (l+m+1)}}P_l^m(\cos\theta)e^{im\varphi}\,.
  \end{align}
With this normalization we conjecture that the transformation between $T^{ll}_i$ and $Y_l^m$ is unitary:
  \begin{align}
    T_i^{ll} = \sum_m c^{llm}_{\ve i} Y_l^m\,,\quad Y_l^m = \sum_{\ve i} \overline{c_{\ve i}^{llm}}T_{\ve i}^{ll}\,.
  \end{align}
We have checked this relation up to $l=8$ by explicitly calculating the $c^{llm}$. The most convenient way to calculate is to express $\hat{\ve r}$ in the complex basis ${\hat{\ve e}_{-1}, \hat{\ve e}_{0}, \hat{\ve e}_{1}}$, related to ${\hat{\ve e}_{x}, \hat{\ve e}_{y}, \hat{\ve e}_{z}}$ by the complex rotation
\begin{align}
\begin{array}{lll}
R_{x, -1}  = \frac{1}{\sqrt 2} &
R_{x, 0}  = 0 &
R_{x, 1}  = -\frac{1}{\sqrt 2} \nn\\
R_{y, -1}  = i \frac{1}{\sqrt 2}& 
R_{y, 0}  = 0 &
R_{y, 1}  = i \frac{1}{\sqrt 2} \nn\\
R_{z, -1}  = 0 &
R_{z, 0}  = 1 &
R_{z, 1}  = 0 \,,
\end{array}
\end{align}
so that 
\begin{align}
  c^{nlm}_{i_1..i_n}=\sum_{\nu_i=-1,0,1}R_{i_1\nu_1}..R_{i_n\nu_n}c^{nlm}_{\nu_1..\nu_n}\,.
\end{align}
In this basis
\begin{align}
  \hat r_\nu=T^{11}_{\nu}&=Y_1^\nu\,,
\end{align}
and therefore
\begin{align}
  \hat r_{\nu_1}..\hat r_{\nu_n} = \sum_{L=0}^{n}\sum_{m=-L}^lc^{nJM}_{\nu_1..\nu_n}Y_J^M &= \sum_{l=0}^{n-1}\sum_{m=-l}^l c^{n-1,lm}_{\nu_1..\nu_{n-1}}Y_l^m Y_1^{\nu_n}\,.
\end{align}
Because of the orthogonality of the spherical harmonics, this leads to a recurrence relation for the coefficients:
\begin{align}
  c^{nJM}_{\nu_1..\nu_n} &= \sum_{l=0}^{n-1}\sum_{m=-l}^l c^{n-1,lm}_{\nu_1..\nu_{n-1}} g(l,m,1,\nu_n,J,M)\,,
\end{align}
where $g$ is the Gaunt coefficient for integrals of the spherical harmonics,
\begin{align}
  g(l_1,m_1,l_2,m_2,J,M) &= \int_S Y_{l_1}^{m_1}Y_{l_2}^{m_2}\overline Y_{J}^{M}\,\rd S\,.
\end{align}

Provided this unitary transformation, we have
\begin{align}
    T^{l_1l_1}_{\ve i}T^{l_2l_2}_{\ve j}=\sum_{J=|l_1-l_2|}^{l_1+l_2}A^{l_1l_1l_2l_2J}_{\ve i\ve j\ve k}T^{JJ}_{\ve k}\,,
  \end{align}
where
  \begin{align}
    A^{l_1 l_1 l_2 l_2 J}_{\ve i\ve j\ve k} &= \sum_{m_1=-l_1}^{l_1} \sum_{m_2=-l_2}^{l_2} \sum_{M=-J}^J c^{l_1 l_1 m_1}_{\ve i}c^{l_2 l_2 m_2}_{\ve j}\overline{c^{J J M}_{\ve k}}g(l_1,m_1,l_2,m_2,J,M)\,.
  \end{align}

\end{document}